\documentclass[twocolumn,aps,showpacs,prb]{revtex4}
\usepackage{color}
\usepackage{graphicx}

\begin{document}
\title{Node-less atomic wave functions, Pauli repulsion and 
  systematic projector augmentation}
\author{Peter E. Bl\"ochl$^{*,1,2}$ and Clemens F\"orst$^{1}$}
\affiliation{$^1$ Clausthal University of Technology, Institute for
Theoretical Physics, Leibnizstr.10, D-38678 Clausthal-Zellerfeld,
Germany}
\affiliation{$^2$ Georg-August Universit\"at G\"ottingen, Institute for
Materials Physics, Friedrich Hund Platz 1, D-37077 G\"ottingen,
Germany}
\date{\today} 
\begin{abstract}
A construction of node-less atomic orbitals and energy-dependent,
node-reduced partial waves is presented, that contains the full
information of the atomic eigenstates and that allows to represent the
scattering properties in a transparent manner.  By inverting the
defining Schr\"odinger equation, the Pauli repulsion by the core
electrons can be represented as effective potential. This construction
also provides a description of the Pauli repulsion by an
environment. Furthermore, the representation leads to a new systematic
scheme for a projector augmentation. The relation to Slater orbitals
is discussed.
\end{abstract}
%
%\keywords{}
\maketitle
%\narrowtext \twocolumn
%=====================================================================
\section{Introduction}
%=====================================================================
First-principles electronic-structure calculations have made a large
contribution to the understanding of materials. Density-functional
theory\cite{hohenberg64_pr136_B864,kohn65_pr140_1133} has proven
itself as an extremely versatile and efficient tool to provide
quantitative information of real materials.  While a number of open
challenges such as strong correlations, embedding of environments,
various property calculations remain, much of the methodology has
matured and has been implemented into efficient program packages.

The development of electronic-structure methods not only has the goal
of \textit{simulating} nature, but also aims to develop an
\textit{efficient language} that allows to describe and rationalize
the properties and the behavior of real materials in simple terms.
This second aspect in turn has been an inspiration for the development
of algorithms that are efficient on the computer.

Both, the pseudopotential approach\cite{topiol77_cpl49_367,
  hamann79_prl43_1494} and the augmented wave methods
\cite{slater37_pr51_846,andersen75_prb12_3060}, which have been linked
by the Projector Augmented Wave (PAW)
method\cite{bloechl94_prb50_17953}, stand in this tradition.  The
pseudopotential approach emphasizes the free-electron-like features,
while the augmented wave methods provide, for example, a sound basis
for a tight-binding description of the electronic
structure\cite{andersen84_prl53_2571}, which is central to the natural
language of chemists.

A central point in discussing the electronic structure is the
separation of the chemical effects of the valence electrons from those
of the core electrons. One can say that augmented wave methods and the
pseudopotential approach came to life by providing their own answer to
this problem.

The goal of this paper is to provide a tool to exploit this separation
of core and valence states from a different point of view. This tool
is a representation of the atomic wave functions in terms of node-less
radial functions. The node-less wave functions provide a description
of the Pauli repulsion by the core electrons by an effective potential
and thus provides a canonical pseudopotential. The construction
described here, is currently used for the pseudization of wave
functions in the PAW method, which is analogous to the construction of
pseudopotentials. It is also used for the natural choice of
tight-binding orbitals, which will be used to incorporate strong
correlations into a DFT environment.

The concept of node-less wave functions have been introduced by
Topiol, Zunger and Ratner\cite{melius74_pra10_1528,
  topiol77_cpl49_367, zunger78_cp30_423, zunger78_prb18_5449} to
construct the first ab-initio pseudopotentials. The resulting
pseudopotentials exhibited a divergent term at the atomic site which
was a disadvantage for plane wave calculations. By relaxing the direct
connection to the atomic wave function, the divergent term could be
avoided, which lead to the norm-conserving
pseudopotentials\cite{hamann79_prl43_1494} used until today.

In this paper, we extend the concept of Zunger et al. by providing a
construction that provides direct contact to the underlying physics of
the atom. We will demonstrate how to exploit this concept for the
construction of the augmentation of the PAW method and how it can be
used to describe the Pauli repulsion to the core states by an
effective semi-local pseudo potential.

%=====================================================================
\section{Definitions and mathematical properties}
%=====================================================================
%=====================================================================
\subsection{Notation}
%=====================================================================
Consider the energy-dependent solution $|\phi(E)\rangle$ of a
spherical Schr\"odinger equation 
\begin{eqnarray}
\biggl(\hat{H}-E\biggr)|\phi(E)\rangle=0
\label{eq:schroedingerpsi}
\end{eqnarray}
with 
\begin{eqnarray}
\lim_{|\vec{r}|\rightarrow0}\frac{\phi(E,\vec{r})}
{r^{\ell}Y_{\ell,m}(\vec{r})}=1\;.
\label{eq:schroedingerpsiinitialcond}
\end{eqnarray}
We suppress the angular-momentum indices $\ell,m$, but assume that the
state $|\phi(E)\rangle$ is an angular-momentum eigenstate,
which can be written as product of a radial function
$R(r,E)$ and a spherical harmonics $Y_{\ell,m}(\vec{r})$, namely as
\begin{eqnarray}
\langle\vec{r}|\phi(E)\rangle=R(|\vec{r}|,E)Y_{\ell,m}(\vec{r})\;.
\label{eq:schroedingerradialspherical}
\end{eqnarray}

When we impose the condition that the wave function vanishes at the
boundary of a sphere with radius $r_x$, we obtain a series of bound
states 
\begin{eqnarray}
|\psi_n\rangle:=|\phi(E_n)\rangle
\end{eqnarray}
with energies $E_n$.

For the time being, we limit the discussion to the non-relativistic
Schr\"odinger equation, i.e.
\begin{eqnarray}
\langle\vec{r'}|\hat{H}|\vec{r}\rangle=\delta(\vec{r}-\vec{r'})
\left[\frac{-\hbar^2}{2m}\vec{\nabla}^2+v(\vec{r})\right]\;.
\end{eqnarray}
Extensions to the Dirac equation will be
presented elsewhere.\cite{schadebloechlpruschke}.

%=====================================================================
\subsection{Construction of node-less bound states}
%=====================================================================
In a first step, we define a discrete series of node-less wave function
$u_n(\vec{r})$ for an atom. 

The series is initiated by the lowest bound state for the specified
angular momentum
\begin{eqnarray}
|u_1\rangle=|\psi_1\rangle\;.
\end{eqnarray}

The series is then constructed recursively by
\begin{eqnarray}
(\hat{H}-E_n)|u_{n}\rangle=-|u_{n-1}\rangle
\label{eq:nodelessbound}
\end{eqnarray}
with the boundary conditions that the functions vanish not only at the
box radius $r_x$, but also at the origin, i.e.
\begin{eqnarray}
\lim_{|\vec{r}|\rightarrow 0}|\vec{r}|^{-\ell-1}\langle\vec{r}|u_n\rangle=0
\qquad \text{for $n>1$}\;.
\end{eqnarray}

As shown below, the energies in this series are the same as the
eigenstates of the Hamiltonian and there is an explicit transformation
between the real wave functions $|\psi_n\rangle$ and the node-less
wave functions $|u_n\rangle$. The transformation has the form
\begin{eqnarray}
|\psi_n\rangle&=&\sum_{m=1}^n |u_m\rangle\prod_{j=1}^{m-1}(E_j-E_n)\;.
\label{eq:fromutopsi}
\end{eqnarray}
The product without a term is defined to be equal to one.  
The identity Eq.~\ref{eq:fromutopsi} can be verified by inserting it into
the Schr\"odinger equation Eq.~\ref{eq:schroedingerpsi} and
the boundary condition Eq.~\ref{eq:schroedingerpsiinitialcond}.

\begin{figure}
\begin{center}
\includegraphics[width=\linewidth,angle=0,clip=true]{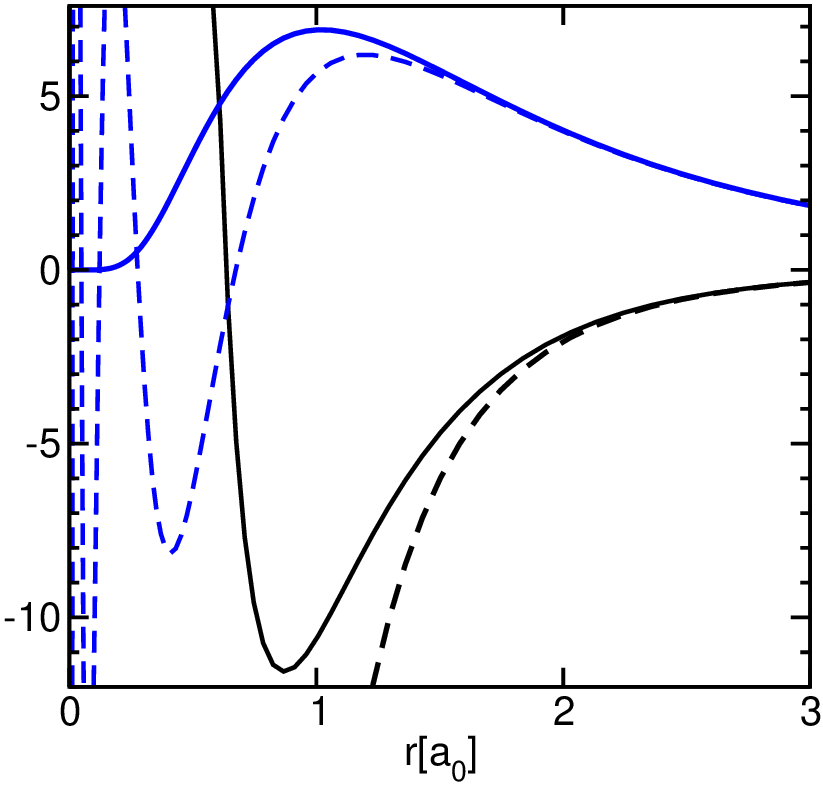}
%{Figs/au6spotandphi.eps}
\end{center}
\caption{\label{fig:au6s}Node-less 6s wave functions shown as full
  lines for Gold (above) and its effective potential (below).  The
  dashed functions are the full wave function and the full
  potential. The scale on the vertical axis is the potential value in
  Hartree.}
\end{figure}

Let us inspect the main properties of the node-less wave
functions. Figure~\ref{fig:au6s} shows the main characteristics for
the 6s wave function of Au in comparison to the normal 6s wave
function.

The radial part of the wave function starts at the origin with a high
power of $|\vec{r}|$. The depression of the wave function near the
nucleus can be loosely attributed to the Pauli repulsion of the core
electrons, that repel the valence electrons from the core region.

The leading power of the node-less wave functions near the origin is
\begin{eqnarray}
\langle\vec{r}|u_n\rangle&=&\left(\frac{2m_e}{\hbar^2}\right)^{n-1}
\frac{(2\ell+1)!!}{(2\ell-1+2n)!! (2n-2)!!}
\nonumber\\ 
&&\hspace{-0.5cm}\times|\vec{r}|^{\ell+2(n-1)}\textrm{e}^{-\lambda_n r}
Y_{\ell,m}(\vec{r})(1+O(|\vec{r}|^2))\;.
\label{eq:leadingpoweru}
\end{eqnarray}
The leading-order term is derived in
appendix~\ref{sec:powerorigin}. The exponential function with the yet
unspecified parameter $\lambda_n$ has been added, because it provides a
good description of the qualitative  behavior of the wave function over a
fairly wide region near the origin.

Eq.~\ref{eq:leadingpoweru} shows that node-reduced partial waves start out
at the origin with a increasingly high power, as $n$ is increased. We
found it surprising that no trace of the nodal structure is left near
the origin.

%=====================================================================
\subsection{Construction of node-reduced wave functions}
%=====================================================================
Now,we introduce energy-dependent partial waves via the
differential equation
\begin{eqnarray}
(H-E)|q_n(E)\rangle=-|u_{n-1}\rangle\;.
\label{eq:defq}
\end{eqnarray}
We will call them \textit{node-reduced partial waves}, because their
number of nodes is $n-1$ less than that of the regular partial wave
$|\phi(E)\rangle$ at the same energy.\footnote{This rule is valid for
  all practical purposes, but it is violated in the scattering
  region.} This construction allows one to effectively remove the core
states from the shape of the wave function and from the scattering
properties.

The boundary conditions at the origin of the node-reduced partial
waves are chosen analogously to those for $|u_n\rangle$, namely
\begin{eqnarray}
\lim_{|\vec{r}|\rightarrow0} \frac{q_n(E,\vec{r})}{u_n(\vec{r})}=1\;.
\label{eq:qboundaryc}
\end{eqnarray}
$|q_1(E)\rangle$ is identical to the regular energy-dependent partial
wave $|\phi(E)\rangle$ with the normalization from
Eq.~\ref{eq:schroedingerpsiinitialcond}.

Using $E=E_n$ and comparing Eq.~\ref{eq:defq} with
Eq.~\ref{eq:nodelessbound}, we obtain that the bound states of the
node-reduced partial waves are the node-less wave functions defined
before, i.e. 
\begin{equation}
|q_n(E_n)\rangle=|u_n\rangle\;.
\label{eq:qofen}
\end{equation}

The energy-dependent wave function $|\phi(E)\rangle$ can be
reconstructed from the node-reduced wave function $|q_n(E)\rangle$ and
the node-less bound states $|u_n\rangle$ by
\begin{eqnarray}
|\phi(E)\rangle&=&
\biggl[|q_n(E)\rangle
+\sum_{m=1}^{n-1}|u_m\rangle
\prod_{j=m}^{n-1}\frac{1}{E_j-E}\biggr]
\nonumber\\
&&\times
\prod_{j=1}^{n-1}(E_j-E)\;.
\label{eq:psiofefromqofe}
\end{eqnarray}
Thus we obtained a partition of the wave function into a
node-reduced part and a part related to the core wave functions, which
is again expressed in terms of node-less wave functions.

As shown in appendix~\ref{sec:lemma1}, the node-reduced wave function
can be expressed as difference 
\begin{eqnarray}
|q_{n+1}(E)\rangle=
\left(|q_{n}(E)\rangle-|q_{n}(E_{n})\rangle\right)
\frac{1}{E-E_{n}}
\label{eq:qsequence}
\end{eqnarray}
of two node-reduced wave functions
$|q_n(E)\rangle$ with one additional node.
According to Eq.~\ref{eq:qsequence}, we can construct all
$|q_n(E)\rangle$ from the energy-dependent partial wave
$|\phi(E)\rangle$, if the latter is normalized such that the
leading term at the origin remains energy-independent.

Specifically for the bound state energies, the node-reduced wave
functions are related, as shown in appendix~\ref{sec:lemma2}, to the
energy derivatives of other node-reduced wave functions.
\begin{equation}
|q_{n+j}(E_n)\rangle=\frac{1}{j!}|q_n^{(j)}(E_n)\rangle
\label{eq:qdersequence}
\end{equation}
where 
\begin{equation}
|q^{(j)}_n(E)\rangle:=\partial_E^j|q_n(E)\rangle
\end{equation}
is the $j$-th energy derivative of $|q_n(E)\rangle$.  

Using Eq.~\ref{eq:qdersequence}, we find that the node-reduced wave
functions are related to each other by a Taylor expansion about one
bound state energy
\begin{eqnarray}
|q_n(E)\rangle=\sum_{j=0}^\infty |q_{n+j}(E_n)\rangle(E-E_n)^j
\end{eqnarray}

%=====================================================================
\subsection{Towards a nodal-theorem for node-reduced wave functions}
%=====================================================================
Eq.~\ref{eq:psiofefromqofe} sheds some light on the question of the
node-less-ness of the wave function $|u_n\rangle$: 

We will show later, that the assumptions of the nodal theorem for the
node-reduced wave functions is violated in the scattering
region. Nevertheless, the assumptions are valid in the energy region
of bound states. It is our hope that a variant of the nodal theorem
can be found that holds for the entire energy region.

A nodal theorem for $|q_n(E)\rangle$ requires one to make the
assumption that the node-less functions $|u_m\rangle$ with $m<n$ do
not reach the outer boundary $|\vec{r}|=r_x$ for the node count.  This
assumption is reasonable for bound states of an atom if the outer
boundary is chosen sufficiently far away.

With this assumption, the decomposition of the wave function in
Eq.~\ref{eq:psiofefromqofe} shows that $|q_n(E)\rangle$ has bound
states for all energies $E_j$ with $j\ge n$.  At the lower bound
states, that is for $j<n$ the node-less wave function
$|q_n(E)\rangle$ does not have a node, but its contribution is removed
by the energy-dependent factors in Eq.~\ref{eq:psiofefromqofe}.

Under the same assumption, the regular nodal
theorem\cite{courant24_book1} shows that nodes of the node-reduced
partial waves at the outer bound move inward. Thus the number of
nodes increases by one at each energy $E_j$ with $j\ge n$.

We can also exclude that nodes leave the interval at the center,
because the leading order of the Taylor expansion about the origin is,
according to Eq.~\ref{eq:qofen}, energy independent. If a node would
migrate across the origin, the leading order of $|q(E)\rangle$ would
change sign in contradiction to Eq.~\ref{eq:qofen}.

Two additional assumptions need to be made, namely that there is no
pair-wise creation or annihilation of nodes within the interval, and
that there are no accidental nodes of the node-reduced wave function
$|q_n(E)\rangle$ at the bound state energies $E_j$ with $j<n$.

\begin{figure}
\begin{center}
\includegraphics[width=\linewidth,angle=0,clip=true]{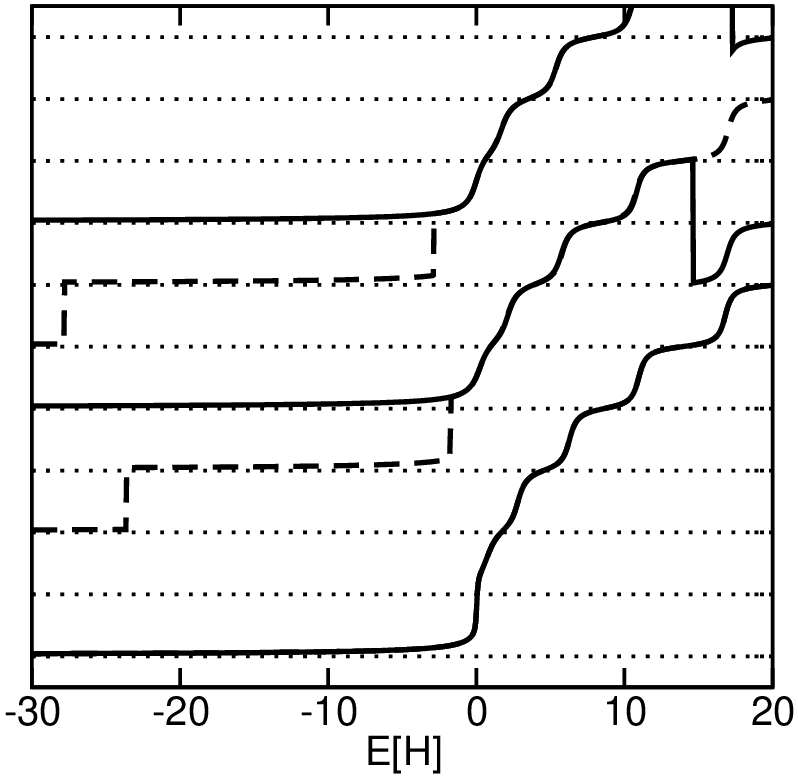}
%{Figs/fescatt.eps}
\end{center}
\caption{\label{fig:fescatt} Phase shift function of Fe as defined by
  $N(E)+\frac{1}{2}-\frac{1}{\pi}\textrm{arctan}(\partial_r\ln(\phi(E,r)))$
  as function of energy. $N(E)$ is the number of nodes within the
  radius at which the derivative of the wave function is taken.  The
  phase shift functions have been displaced individually by integers
  for better comparison.  The dotted lines mark integer values.  The
  full lines are the phase shifts of the node-reduced partial wave and
  the dashed lines are those partial waves with the original nodal
  structure. From top to bottom, the s,p,d angular momentum channels
  are shown. The energy region between half-integer and the next
  integer value is the band region for this angular momentum
  channel. The steps of the dashed lines between -30 and -20~H reflect
  the 2s,2p shell and the one shortly below zero is the 3s,3p
  shell. The jumps in the high energy region reflect the annihilation
  of two nodes within the augmentation region.}
\end{figure}

Thus, for reasonable assumptions, these arguments show that the bound
wave functions are free of nodes. In the scattering region at high
energies though, the assumptions made are violated.  Numerical
calculations show violations of this nodal theorem, namely the
pairwise formation of two nodes, in the high-energy region, which can
be seen in Fig.~\ref{fig:fescatt}.

As a consequence of the nodal theorem presented here, the number of
nodes of the node-reduced wave function $|q_n(E)$ differs from that of
the full wave functions by $n-1$ for energies above $E_n$ and it
vanishes below.

%=====================================================================
\subsection{Properties}
%=====================================================================

%=====================================================================
\subsubsection{Slater orbitals}
%=====================================================================
In 1929 Slater suggested a simple form for node-less atomic
orbitals\cite{slater30_pr36_57} of the form
\begin{eqnarray}
\chi(\vec{r})=A r^{n^*-1}\mathrm{e}^{-\frac{Z-s}{n^* a_0} r}\;,
\end{eqnarray}
where $n^*$ is an effective main quantum number, Z is the atomic
number of the atom, $s$ is a screening constant, $a_0$ is the Bohr
radius and $A$ is a normalization constant.

The guiding idea of Slater is related to ours, namely to exploit that
the number of nodes is of only secondary importance for many of the
chemical properties, if only the outer tail of the wave function is
properly described. In this regard, our node-less functions can be
considered an extension of Slater's orbitals. In contrast to Slater's
orbitals our node-less functions contain, as a set, the complete
information of the wave functions, at the cost that they do not have
the simple analytical form of Slater's orbitals.

\begin{figure}
\begin{center}
\includegraphics[width=\linewidth,angle=0,clip=true]{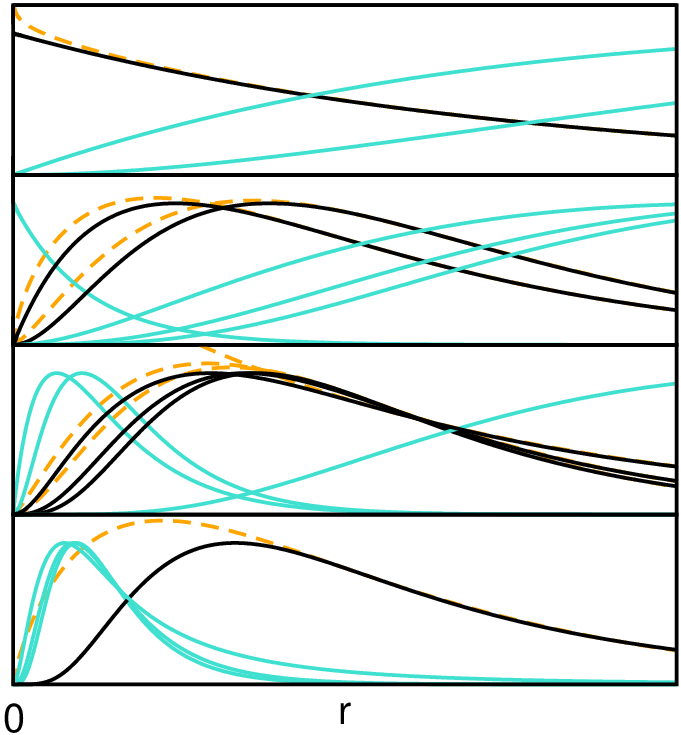}
%{Figs/feushells.eps}
\end{center}
\caption{\label{fig:feushells}Node-less wave functions of the Fe atom
  and comparison with Slater orbitals as function of the distance from
  the center. Each graph shows in black the orbitals from one main
  quantum number (full black line) and the corresponding Slater
  orbital (orange dashed line). The orbitals of the neighboring main
  quantum numbers are show in (green, full lines). The wave functions
  are scaled to the same maximum value.  The horizontal axis is scaled
  for each graph individually to make the shape similarity evident.
  The range of radii on the horizontal axis extend from the origin to
  0.05, 0.35, 1.3, and 5~a$_0$, respectively from top to bottom.}
\end{figure}

The qualitative behavior and the similarity to Slater's orbitals have
been investigated by first by Thieme\cite{thieme03_diploma} and
F\"orst\cite{foerst04_thesis}. The series of node-less functions
is shown in Fig.~\ref{fig:feushells}.

\begin{figure}
\begin{center}
\includegraphics[width=\linewidth,angle=0,clip=true]{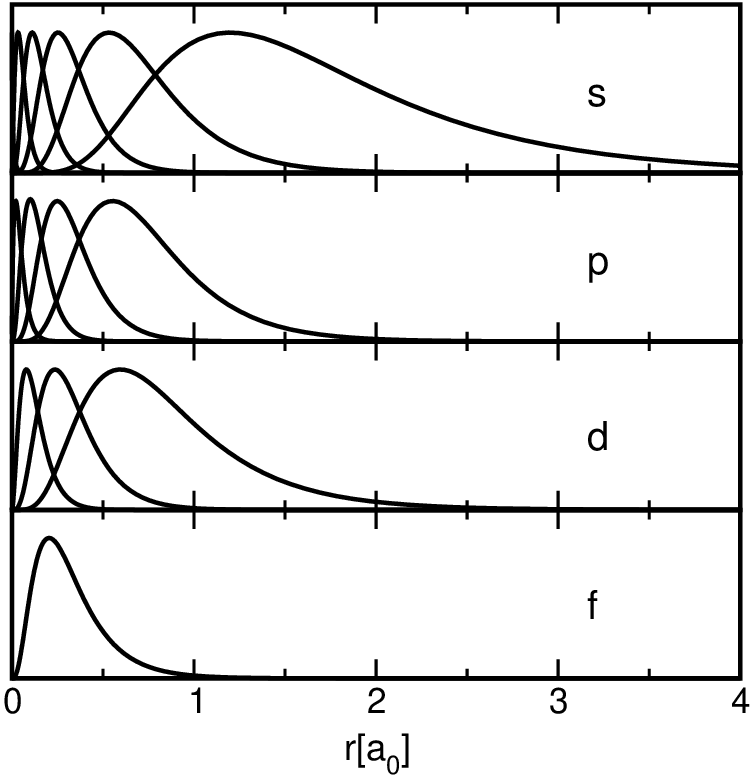}
%{Figs/nodlessphiau.eps}
\end{center}
\caption{\label{fig:nodelessgoldandexponential} Node-less wave
  functions for Gold for s,p,d and f electrons from top to bottom. The
  wave functions are normalized to have identical maximum value}
\end{figure}

Interesting to see is that the node-less 3d wave function of Fe is
much closer in space to the 3s and 3p node-less wave functions than to
the 4s wave function, to which it is energetically closer. It has been
a general observation that the the node-less wave functions group
according to their main quantum number into sets with similar range,
and that these sets are well separated from those with a different main
quantum number. With increasing atomic number, the wave functions with
same main quantum number become even more similar as seen in
Fig.~\ref{fig:nodelessgoldandexponential}.

The Slater orbitals, shown as dashed lines in
Fig.~\ref{fig:feushells}, have been parameterized to reproduce the
radial density maximum, i.e. the maximum of $r^2u_n^2(r)$, in value
and position and the logarithmic derivative at a distance 1.5 times
that of the density maximum. The tails of the wave function match
fairly well, while there are deviations in the inner region. Compared
to the Slater orbitals, which decay with a power time an exponential,
the node-less wave functions have a simple exponential tail.

%=====================================================================
\subsubsection{Pauli repulsion}
%=====================================================================
\begin{figure}
\begin{center}
\includegraphics[width=7cm,angle=0,clip=true]{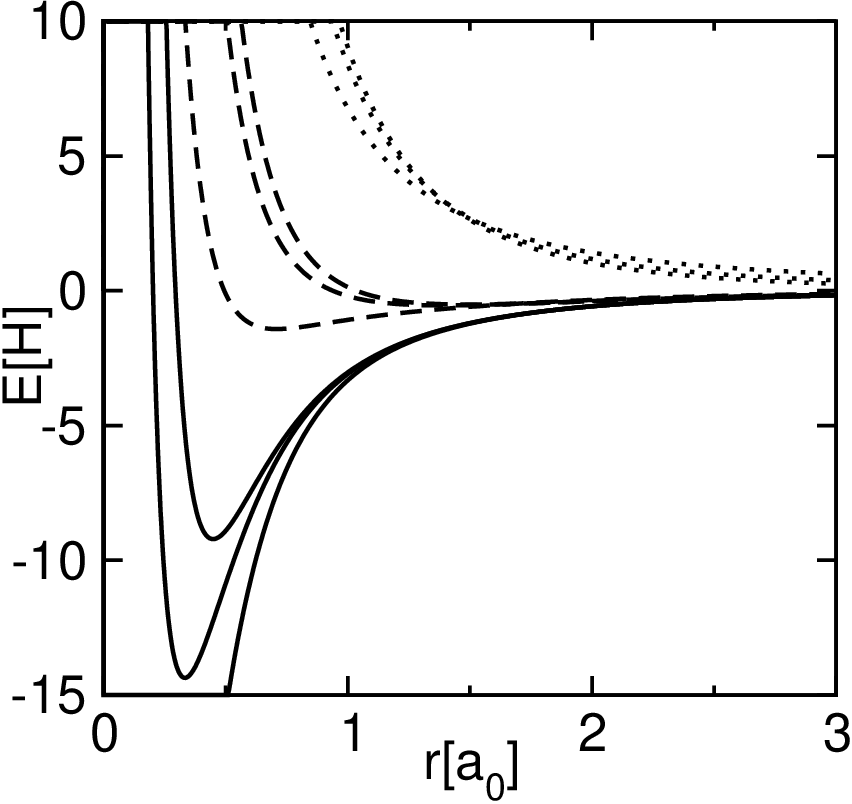}
%{Figs/fevpauli.eps}
\end{center}
\caption{\label{fig:fevpauli} Effective potential including Pauli
  repulsion for the 3s3p3d shell (full) the 4s4p4d shell (dashed) and
  the 5s5p5d shell (dotted) of Fes$^2$d$^6$. For the shell with main
  quantum number 3 and 4, the effective potentials decrease with
  increasing angular momentum. For the 3d channel the effective
  potential equals the true potential.}
\end{figure}

One motivation for the node-less construction was to construct
parameter-free pseudopotentials to be used for example to embed a
quantum mechanical calculation into an environment described by a
simpler theory. By inverting the Schr\"odinger equation for the
node-less wave functions, one can construct an effective potential,
that includes the effect of the Pauli repulsion. For a node-reduced
wave function, which obeys Eq.~\ref{eq:defq}, we obtain the potential
with the Pauli repulsion by
\begin{eqnarray}
v_{eff,n}(\vec{r})=E
+\frac{\hbar^2\vec{\nabla}^2u_n(\vec{r})}{2m u_n(\vec{r})}\;.
\end{eqnarray}
The potential is semi-local, that is a different potential acts on
each angular momentum channel. Here, only a specific angular momentum
channel is considered.

The effective potential can also be constructed from the node-reduced
partial waves, $|q_n(E)\rangle$, but this requires an energy below the
first bound state to avoid poles in the potential resulting from the
nodes in the wave function. The energy derivative of this potential
provides an overlap operator, that, however, can also be obtained
directly from the norm of the node-less wave functions as compared to
the true nodal wave function.

In Fig.~\ref{fig:fevpauli}, the effective potentials are shown for the
shells with main quantum number 3 to 5 for iron. The potentials become
more repulsive with increasing angular momentum. 

Interestingly, the potentials form groups with similar behavior
characterized by the main quantum number. This implies that a
semilocal potential constructed from node-less wave functions with the
same main quantum number is best approximated by a local
potential. Such choices will also lead to the most stable and
transferable pseudopotentials:  In the case of Fe, shown in the example,
it is advisable to include the 3s and 3p electrons as valence
electrons.

Atomic wave functions without nodes, such as 3d wave functions, are
special, because they do not experience any Pauli repulsion, so that
the effective potential is equal to the total potential. For atoms
having wave functions without nodes in the valence shell, such as
first-row elements, 3d transition-metal elements or lanthanides, the
intrinsic non-locality as judged from our Pauli repulsion potentials
is strongest. It is these elements for which the construction of
pseudopotentials is most difficult.

Currently, we are using these effective potentials to construct
tight-binding orbitals which will be discussed in a forthcoming
paper\cite{bloechl_ntbo}.

%=====================================================================
\subsubsection{Scattering properties}
%=====================================================================
The shape of the node-less functions makes them suitable helper
functions for the construction of pseudo wave functions for the PAW
method. 

Pseudo wave functions shall be identical to the true wave functions in
the bonding region.  The s- and p- wave function of transition metal
ions have their outmost nodes fairly close to the covalent radius. As
a result a pseudo wave function follows the all electron wave function
into the downswing towards the outermost node. The requirement of
node-less-ness for the pseudo wave function requires a sharp turn
outside the node and a large region with low amplitude. This sharp
turn imposes large amplitudes in the pseudopotential causing stability
and transferability problems. If the construction is such that the
pseudo wave function follows the node-less wave function this problem
is largely avoided.

A second difficulty of constructing pseudo wave function is to avoid
artificial core-like states. They can be induced by the proximity to
semi-core states that are excluded from the valence shell. Because the
pole of the logarithmic wave function of these states may be
noticeable, the pseudized wave functions try to reproduce this pole
resulting in artificial undesired states. When the node-reduced wave
functions are used as basis for the construction of the
pseudo-potential, this problem is avoided.

%=====================================================================
\subsubsection{Towards a systematic projector augmentation}
%=====================================================================
In the PAW method\cite{bloechl94_prb50_17953}, the true Kohn-Sham wave
functions $|\Psi\rangle$ are constructed as
\begin{eqnarray}
|\Psi\rangle=|\tilde{\Psi}\rangle
+\sum_\alpha\Bigl(|\varphi_\alpha\rangle-|\tilde{\varphi}_\alpha\rangle\Biggr)
\langle\tilde{p}_\alpha|\tilde\Psi\rangle
\end{eqnarray}
from auxiliary wave functions $|\tilde\Psi\rangle$ by augmenting them
with the difference of all-electron and auxiliary partial waves,
$|\phi_\alpha\rangle$ and $|\tilde{\phi}_\alpha\rangle$
respectively. The coefficients are obtained using the projector
functions $\langle\tilde{p}_\alpha|$, that obey a biorthogonality
condition
\begin{eqnarray}
\langle\tilde{p}_\alpha|\tilde{\phi}_\beta\rangle=\delta_{\alpha,\beta}
\label{eq:biortho}
\end{eqnarray}
with the auxiliary partial waves.

Node-less wave functions guide us to a more systematic construction of
the augmentation of the projector augmented wave method. So far, the
mapping between all-electron and pseudo partial waves has been done
for a selected grid of energies\cite{bloechl94_prb50_17953}. One
difficulty is to ensure that the choice of auxiliary partial waves is
consistent, that is to ensure that the augmentation leads to a
converging series.

By inspecting some of the conventional pseudization schemes, we
observed that the difference in shape between the resulting
energy-dependent auxiliary partial waves $|\tilde{\phi}_\alpha\rangle$
and the node-reduced wave functions $|q_n(E_\alpha)\rangle$ depends
only weakly on energy. This suggests a new approach for choosing the
projector augmentation.

We propose the following ansatz for the auxiliary partial waves,
namely
\begin{eqnarray}
|\tilde{\phi}(E)\rangle=|q_n(E)\rangle+|k\rangle\;,
\label{eq:ansatzphi}
\end{eqnarray}
where $|k\rangle$ is energy-independent function that is entirely
localized in the augmentation region. A convenient way to choose
$|k\rangle$ is to construct the pseudo partial wave with a
conventional method for one single energy for each angular momentum
and then to resolve Eq.~\ref{eq:ansatzphi} for $|k\rangle$.

Following the recipe of the PAW method\cite{bloechl94_prb50_17953},
one constructs raw, energy dependent projector functions via
\begin{eqnarray}
|\tilde{p}'(E)\rangle=
\biggl(\frac{\hat{\vec{p}}^2}{2m_e}+\tilde{v}-E\biggr)
|\tilde{\phi}(E)\rangle\;.
\label{eq:bareprojector}
\end{eqnarray}
The auxiliary potential $\tilde{v}$ is a local potential, that is
identical to the true potential beyond the augmentation region and
smooth inside. The prime has been attached to the raw projector
functions to distinguish them from the final projector functions
obtained by enforcing the biorthogonalization Eq.~\ref{eq:biortho}.

Insertion of the Ansatz Eq.~\ref{eq:ansatzphi} into
Eq.~\ref{eq:bareprojector} results, with the help of
Eq.~\ref{eq:defq}, in
\begin{eqnarray}
|\tilde{p}'(E)\rangle&=&
-|u_{n-1}\rangle
+\left(\tilde{v}(\hat{\vec{r}})-v(\hat{\vec{r}})\right)|q_n(E)\rangle
\nonumber\\
&+&\biggl(\frac{\hat{\vec{p}}^2}{2m_e}+\tilde{v}-E\biggr)|k\rangle\;.
\end{eqnarray}

Instead of using a grid of energies we can use the Taylor expansion of
this expression in the energy about some value $E_\nu$.
\begin{eqnarray}
|\tilde{p}^{\prime(j)}(E_\nu)\rangle&=&
\left[-|u_{n-1}\rangle
+\biggl(\frac{\hat{\vec{p}}^2}{2m_e}+\tilde{v}-E_\nu\biggr)
|k\rangle\right]\delta_{j,0}
\nonumber\\
&&-|k\rangle\delta_{j,1}
+\left(\tilde{v}(\hat{\vec{r}})-v(\hat{\vec{r}})\right)
|q_n^{(j)}(E_\nu)\rangle\;,
\end{eqnarray}
where $|\tilde{p}^{\prime(j)}(E_\nu)\rangle$ is the $j$-th energy
derivative of $|\tilde{p}^{\prime}(E)\rangle$ at the energy $E_\nu$.
Note, that the last term vanishes quickly with increasing $j$, because
the difference of the potentials is concentrated in the inner region
of the atom, while the radius, beyond which $|q^{(j)}_n\rangle$
becomes appreciable, rapidly shifts further out with each increasing
$j$.

The Taylor expansion of the auxiliary partial waves defined in
Eq.~\ref{eq:ansatzphi} is 
\begin{eqnarray}
|\tilde{\phi}^{(j)}(E_\nu)\rangle=|q_n^{(j)}(E_\nu)\rangle
+|k\rangle\delta_{j,0}\;.
\end{eqnarray}

Finally, the biorthogonality condition is enforced either by a
Gram-Schmidt like bi-orthogonalization or by inversion.
\begin{eqnarray}
|\tilde{p}_j\rangle&=&\sum_{j'}|\tilde{p}^{\prime(j')}(E_\nu)\rangle
\biggl(\langle\tilde{\phi}^{(j)}(E_\nu)
|\tilde{p}^{\prime(j)}(E_\nu)\rangle\biggr)^{-1}
_{j',j}
\nonumber\\
|\tilde{\varphi}_j\rangle&=&|\tilde{\phi}^{(j)}(E_\nu)\rangle
\nonumber\\
|\varphi_j\rangle&=&|\phi^{(j)}(E_\nu)\rangle\;.
\end{eqnarray}
Tests for this class of projector augmentation will be presented in a
forthcoming paper.

%=====================================================================
\section{Conclusions}
%=====================================================================
A set of functions has been presented that provides the description of
the atomic bound and scattering wave functions without the obscuring
nodal structure. An explicit transformation mediates between the
original wave functions and what we call ``node-less'' or
``node-reduced'' wave functions.

This set of wave functions finds applications in the construction of
pseudopotentials or the auxiliary partial waves in the PAW method.  It
provides an effective potential that describes the Pauli-repulsion of
the core electrons without explicit orthogonalization.  The node-less
wave functions will be used to define tight-binding orbitals for
the analysis of chemical binding or the calculation of correlated
materials.

\textbf{Acknowledgement:} Financial support by the Deutsche
Forschungsgemeinschaft through FOR 1346 is gratefully acknowledged.

\appendix
%=====================================================================
\section{Power-series expansion at the origin}
\label{sec:powerorigin}
%=====================================================================
The behavior of the node-reduced wave functions at the origin is
analyzed by the power-series expansion of the corresponding
inhomogeneous differential equation Eq.~\ref{eq:defq}.  The radial part of the
node-reduced wave function is represented as $q(E,r)=\sum_j
a_jr^{\ell+j}$, that of the inhomogeneity as
$u_{n-1}(r)=-\sum_jg_jr^{\ell+j}$ and the potential is expressed as
\begin{eqnarray}
v(r)=-\frac{Ze^2}{4\pi\epsilon_0r}+\sum_{m=0}^\infty v_m r^m
\end{eqnarray}
The radial part of the differential equation Eq.~\ref{eq:defq} has the form
\begin{eqnarray}
&&\Bigl[\frac{-\hbar^2}{2m_e}
\left(\frac{1}{r}\partial^2_r r-\frac{\ell(\ell+1)}{r^2}\right)
\nonumber\\
&&-\frac{Ze^2}{4\pi\epsilon_0r}+\sum_{m=0}^\infty v_m r^m-E\Bigr]
\sum_{j} a_{j} r^{\ell+j}
=\sum_j g_j r^{\ell+j}
\nonumber
\end{eqnarray}
It results in the following recursion for the coefficients $a_j$.
\begin{eqnarray}
a_{j}
&=&
\frac{2m_e}{-\hbar^2(2\ell+j+1)j}
\biggl[\frac{Ze^2}{4\pi\epsilon_0} a_{j-1} 
\nonumber\\
&&-\sum_{m=0}^\infty a_{j-m-2} (v_m-E\delta_{m,0}) 
+ g_{j-2} \biggr]
\end{eqnarray}

The regular solution of the homogeneous problem starts with $a_{0}$,
i.e. $a_j=0$ for $j<0$.  The irregular solution of the homogeneous
problem starts with $a_{-2\ell-1}$, i.e. $a_j=0$ for $j<-2\ell-1$.
For the inhomogeneous differential equation, we can add a solution of
the homogenous problem, so that the leading term of the solution is
determined by the inhomogeneity, given that the leading power of the
inhomogeneity is higher than $r^\ell$.  If the lowest-order term of the
inhomogeneity is $g_m$, the lowest-order term of the wave function is
of order $m+2$, i.e. $a_j=0$ for $j<m+2$.
\begin{eqnarray}
a_{j}=-\frac{2m_e}{\hbar^2(2\ell+j+1)j} g_{j-2} 
\end{eqnarray}

Thus the behavior at the origin is
\begin{eqnarray}
\langle\vec{r}|q_n(E)\rangle&=&\left(\frac{2m_e}{\hbar^2}\right)^{n-1}
\frac{(2\ell+1)!!}{(2\ell-1+2n)!! (2n-2)!!}
\nonumber\\
&&\times|\vec{r}|^{\ell+2(n-1)}Y_{\ell,m}(\vec{r})(1+O(|\vec{r}|))
\label{eq:leadingpowerqapp}
\end{eqnarray}
Thus the leading power increases by two with increasing main quantum
number, which describes the repulsion from the origin.  The prefactor
of the leading power is furthermore energy-independent.

%===================================================================
\section{Recursion for energy dependent node-reduced wave functions}
\label{sec:lemma1}
%===================================================================
Here we show that $|q_n(E)\rangle$ for $n>1$ can be represented as
difference between two functions $|q_{n-1}(E)\rangle$ at different
energies by Eq.~\ref{eq:qsequence}.

We start with a trial function $|f(E)\rangle$ defined by
Eq.~\ref{eq:qsequence}:
\begin{eqnarray}
|f(E)\rangle:=
\Bigl(|q_{n-1}(E)\rangle-|q_{n-1}(E_{n-1})\rangle\Bigr)
\frac{1}{E-E_{n-1}}
\label{eq:deff_trial}
\end{eqnarray}
and show that it satisfies all conditions for $|q_n(E)\rangle$, namely
differential equation Eqs.~\ref{eq:defq} and boundary conditions
Eq.~\ref{eq:qboundaryc}.

Using Eq.~\ref{eq:deff_trial}, together with Eqs.~\ref{eq:defq} and
\ref{eq:qofen} for the node-reduced functions $|q_{n-1}(E)\rangle$, we
find
\begin{eqnarray}
(H-E)|f(E)\rangle&=&|u_{n-1}\rangle
\end{eqnarray}
which is identical to the differential equation
Eq.~\ref{eq:defq} for $|q_n(E)\rangle$.

The boundary conditions at the origin are defined by the admixture of
the homogeneous solutions, that have the leading power $-\ell-1$ and
$\ell$. Because the leading power of $|q_{n-1}(E)\rangle$ in $r$ is,
due to Eq.~\ref{eq:leadingpowerqapp}, energy independent, and higher
or equal to $r^\ell$, the leading power of $|f(E)\rangle$ is
determined by the inhomogeneity. Thus the boundary behavior is
determined directly by the differential equation, which is identical
for $|f(E)\rangle$ and $|q_n(E)\rangle$

Since $|f(E)\rangle$ and $|q_n(E)\rangle$ obey
the same inhomogeneous differential equation and the same boundary
conditions, namely value and derivative at the origin, they are
identical, which proves Eq.~\ref{eq:qsequence}.

%===================================================================
\section{Energy derivatives of node-reduced wave functions}
\label{sec:lemma2}
%===================================================================
Here, we show that the energy-derivatives of the node-reduced wave
functions at the energy $E_n$ of a specific bound state are related by
Eq.~\ref{eq:qdersequence}, namely
\begin{eqnarray}
|q_{n+j}(E_n)\rangle=\frac{1}{j!}|q^{(j)}_{n}(E_n)\rangle
\label{eq:qdersequencelemma2}
\end{eqnarray}
where the $q^{(j)}_n(E)$ is the $j$-th energy derivative of $|q_n(E)\rangle$.

We start with Eq.~\ref{eq:qsequence} for $|q_{n+1}(E)\rangle$
and insert the Taylor expansion for $|q_n(E)\rangle$ about $E_n$
\begin{eqnarray}
|q_{n+1}(E)\rangle&\stackrel{Eq.~\ref{eq:qsequence}}{=}&
\frac{|q_n(E)\rangle-|q_n(E_n)\rangle}{E-E_n}
\nonumber\\
%&=&\stackrel{\textrm{Taylor in }E}{=}
%&&\hspace{-1cm}=
%\frac{1}{E-E_n}
%\left[\left(\sum_{j=0}^\infty \frac{1}{j!}(E-E_n)^j q^{(j)}_n(E_n)\right)-q_n(E_n)\right]
%\nonumber\\
&=&\sum_{j=1}^\infty \frac{1}{j!}|q^{(j)}_n(E_n)\rangle(E-E_n)^{j-1}
\end{eqnarray}
Term-by-term comparison with the Taylor expansion of
$|q_{n+1}(E)\rangle$ on the left-hand side yields
\begin{eqnarray}
|q^{(j)}_{n+1}(E_n)\rangle&=&\frac{1}{j+1}|q^{(j+1)}_n(E_n)\rangle
\label{eq:qdersequencederivationeq3}
\end{eqnarray}

Successive application of Eq.~\ref{eq:qdersequencederivationeq3} leads
to the desired result, Eq.~\ref{eq:qdersequencelemma2}
respectively, Eq.~\ref{eq:qdersequence}.

\end{document}